\documentclass[aps,pra,twoside,twocolumn,a4paper,reprint,noeprint]{revtex4-2}
\usepackage[utf8]{inputenc}
\usepackage{graphicx}
\usepackage{amsmath}
\usepackage{amsfonts}
\usepackage{amssymb}
\usepackage{mathtools}
\usepackage[usenames,dvipsnames,svgnames]{xcolor}
\usepackage[USenglish]{babel}
\usepackage{hyperref}
\usepackage{grffile}
\usepackage{caption}
\usepackage{subcaption}
\usepackage{booktabs}
\usepackage{rotating}
\usepackage[all]{nowidow}

\allowdisplaybreaks
\begin{document}

\title{Majorana flat bands at structured surfaces of nodal noncentrosymmetric superconductors}

\author{Clara J. Lapp}
\email{clara{\_}johanna.lapp@tu-dresden.de}
\affiliation{Institute of Theoretical Physics, Technische Universit\"at Dresden, 01069 Dresden, Germany}
\affiliation{W\"urzburg--Dresden Cluster of Excellence ct.qmat, Technische Universit\"at Dresden, 01062 Dresden, Germany}

\author{Carsten Timm}
\email{carsten.timm@tu-dresden.de}
\affiliation{Institute of Theoretical Physics, Technische Universit\"at Dresden, 01069 Dresden, Germany}
\affiliation{W\"urzburg--Dresden Cluster of Excellence ct.qmat, Technische Universit\"at Dresden, 01062 Dresden, Germany}

\date{\today}

\begin{abstract}
Surfaces of nodal noncentrosymmetric superconductors can host flat bands of Majorana modes, which provide a promising platform for quantum computation if one can find methods for manipulating localized Majorana wave packets. We study the fate of such flat bands when part of the surface is subjected to an exchange field induced by a ferromagnetic insulator. We use exact diagonalization to find the eigenstates and eigenenergies of the Bogoliubov--de Gennes Hamiltonian of a model system, for which an exchange field is applied along a strip on the surface of a slab. We consider different orientations of the strip and the applied field. If the spin polarization of the field-free system along the field direction is sufficiently large perturbation theory predicts that energies of states which are mostly localized on the exchange-field strip are shifted away from zero energy by an amount proportional to the field strength. On the other hand, energies corresponding to states localized on the field-free strip are only weakly affected by the field. Exact diagonalization confirms this. Moreover, we discuss a setup with a small exchange field applied to the previously field-free strip with the goal of introducing a linear dispersion. By switching this dispersion on and off, a wave packet could be moved in a certain direction. We find that in our model system, a linear dispersion can indeed be achieved. The qualitative features of this dispersion can be predicted from the momentum-dependent spin polarization of the field-free surface.
\end{abstract}

\maketitle

\section{Introduction}
\label{sec:intro}

For noncentrosymmetric superconductors (NCSs) that obey time-reversal symmetry and exhibit line nodes of the superconducting gap function, one can define a momentum-dependent invariant that ensures the existence of flat bands of zero-energy surface states in regions of the surface Brillouin zone where this invariant is nonzero~\cite{SR11,BST11,SBT12,STY11,HQS13,SB15}. Due to being their own antiparticles, these surface modes are also known as Majorana modes.

As the flat surface bands occupy a nonzero fraction $S_f/S_{\text{BZ}}$ of the area $S_{\text{BZ}}$ of the surface Brillouin zone, it is possible to construct linear combinations of them which are localized at arbitrary points in real space~\cite{RRT20}. However, the number of independent localized zero-energy modes in real space is not equal to the number of sites at the surface, i.e., the number of points in the surface Brillouin zone, but is reduced by a factor of $S_f/S_{\text{BZ}}$. For the same reason, the wave packets have a minimal width in real space that is inversely proportional to the maximal diameter of the support of the flat bands in the surface Brillouin zone. Moreover, these wave packets have zero eigenenergy only in the limit of an infinitely thick slab.

Majorana modes have attracted a lot of interest in the context of quantum computation \cite{Kit03,NSSFS08,SLTD10,LSD10,ORO10,ElF15,SFN15,OrO20}. Such applications require ways to move Majorana modes around and, in particular, to move them past each other so as to braid their world lines. The behavior upon braiding is of physical relevance because quasiparticles in a two-dimensional system can display anyon statistics. These anyons are called Abelian if, upon exchanging, they can gather any phase factor $e^{i\phi}$. On the other hand, for non-Abelian anyons, the braiding operations do not commute anymore. Localized Majorana modes at the surfaces of NCSs provide a promising platform for this. Here, we make progress in two ways: First, it is necessary to confine the Majorana modes to a certain real-space region of the surface, i.e., to construct Majorana circuits. Second, one also has to be able to move Majorana wave packets, which requires a time-dependent modification of the bands such that they are weakly dispersing.

In this paper, we suggest one introduce time-reversal-symmetry-breaking terms to the Hamiltonian to achieve both objectives. These terms are realized by exchange fields applied to parts of the surface by bringing the superconductor into contact with a ferromagnetic insulator.

If an exchange field is applied to the entire surface, this leads to a tilting of the previously flat surface bands away from zero energy~\cite{BTS13,STB13} due to the momentum-dependent spin polarization of the surface states~\cite{BST15}. We first use this effect to restrict the zero-energy surface states to certain strips on the surface by applying a strong exchange field everywhere else. For a nonzero spin polarization of the surface states in the field-free system, this generically leads to the localization of low-energy surface modes in the field-free strip. We then introduce a small exchange field to this previously field-free strip to induce a weak dispersion in order to move a Majorana wave packet along the strip.

The remainder of this paper is organized as follows. In Sec.\ \ref{sec:Model system}, we introduce the model used for our analysis. In Sec.\ \ref{sec:Eigenvalues and eigenstates of the Hamiltonian}, we derive and discuss the surface states in the presence of an exchange field applied to strips of various forms. This is followed by an analysis of the prospects of creating a linear dispersion along a strip and thereby moving Majorana wave packets. We summarize our results and draw conclusions in Sec.~\ref{sec:summary}.

\section{Model system}
\label{sec:Model system}

For our calculations, we use a model system with point group $C_{4v}$, which is relevant for, e.g.,  CePt$_3$Si~\cite{BHMPSGSNSR04}, CeRhSi$_3$~\cite{KISUAT05}, CeIrSi$_3$~\cite{SOSYTYMHTSO06}, and LaAlO$_3$/SrTiO$_3$ heterostructures \cite{RTC07}. We determine the low-energy eigenstates and eigenvalues of the Bogoliubov--de Gennes (BdG) Hamiltonian of a $(101)$ slab by a Fourier transformation to real space along the two axes which are not translationally invariant followed by exact diagonalization of the resulting matrix.

\subsection{Hamiltonian}

We start by considering a three-dimensional single band NCS described by the Hamiltonian
\begin{equation}
H = \frac{1}{2} \sum_{\mathbf{k}}\Psi_\mathbf{k}^\dagger \mathcal{H}_\text{BdG}(\mathbf{k})\Psi_{\mathbf{k}} ,
\end{equation}
with the Nambu spinor $\Psi_\mathbf{k} = (c_{\mathbf{k},\uparrow},c_{\mathbf{k},\downarrow},c_{ -\mathbf{k},\uparrow}^\dagger,c_{-\mathbf{k},\downarrow}^\dagger)^\top$ of the electronic creation and annihilation operators for momentum $\mathbf{k}$ and spin $\sigma\in\lbrace\uparrow, \downarrow \rbrace$ and the BdG Hamiltonian
\begin{equation} \label{eq:H_BdG(k)}
\mathcal{H}_{\text{BdG}}(\mathbf{k}) =
\begin{pmatrix}
\epsilon(\mathbf{k}) \hat{\sigma}^0 +\lambda \mathbf{l}_\mathbf{k}\cdot \boldsymbol{\hat{\sigma}}&\hat{\Delta}(\mathbf{k})\\
\hat{\Delta}^\dagger(\mathbf{k})&-\epsilon(\mathbf{k}) \hat{\sigma}^0 + \lambda \mathbf{l}_\mathbf{k}\cdot \boldsymbol{\hat{\sigma}}^*
\end{pmatrix}.
\end{equation}
Here, the vector of Pauli matrices and the $2\times2$ identity matrix are denoted by $\boldsymbol{\hat{\sigma}}$ and $\hat{\sigma}^0$, respectively, and
\begin{equation} \label{eq:h(k)}
h(\mathbf{k}) = \epsilon(\mathbf{k}) \hat{\sigma}^0 + \lambda \mathbf{l}_\mathbf{k}\cdot \boldsymbol{\hat{\sigma}}
\end{equation}
is the normal-state Hamiltonian. The first term, $\epsilon(\mathbf{k}) \hat{\sigma}^0$, which is diagonal in the spin basis, will henceforth be represented by the tight-binding dispersion
\begin{equation}
\epsilon(\mathbf{k})=-2t(\cos k_x+\cos k_y +\cos k_z)-\mu
\end{equation}
for nearest-neighbor hopping strength $t$ and chemical potential $\mu$. The second term, $\lambda \mathbf{l}_\mathbf{k}\cdot \boldsymbol{\hat{\sigma}}$, is the antisymmetric spin-orbit-coupling (ASOC) term, in which $\lambda$ represents the spin-orbit-coupling (SOC) strength, while the form of the SOC vector~$\mathbf{l_k}$ is constrained by the lattice symmetries~\cite{SBT12}. For the point group $C_{4v}$, a first-order expansion leads to a Rashba-type SOC with~\cite{S09}
\begin{equation}
\mathbf{l_k} = \mathbf{\hat{x}} \sin k_y
  - \mathbf{\hat{y}} \sin k_x .
\end{equation}

In the energetically most favorable pairing state, the vector of triplet pairing amplitudes tends to be parallel to the ASOC vector $\mathbf{l_k}$~\cite{FAKS04} so that the pairing matrix can be written as
\begin{equation}\label{eq:pairing}
\hat{\Delta}=(\Delta^s \hat{\sigma}^0+\Delta^t \mathbf{l_k}\cdot\boldsymbol{\hat{\sigma}})(i\hat{\sigma}^y),
\end{equation}
with the singlet and triplet pairing strengths $\Delta^s$ and $\Delta^t$, respectively, which we assume to be both constant and positive. Here, $i\hat\sigma^y$ represents the unitary part of the antiunitary time-reversal operation. In this paper, we consider $(s+p)$-wave pairing, as described by Eq.\ \eqref{eq:pairing}, because this leads to the simplest nodal structure. A more complicated momentum dependence of the pairing matrix, e.g., $d$-wave or $f$-wave pairing, would lead to additional nodes and more complex shapes of the momentum-space regions hosting zero-energy surface states \cite{SBT12}. This would increase the computational effort without leading to qualitatively new physical effects.

Diagonalizing Eq.~\eqref{eq:h(k)} leads to two helicity bands
\begin{equation}
\xi^\pm_\mathbf{k}=\epsilon_\mathbf{k}\pm\lambda|\mathbf{l_k}|,
\end{equation}
with the gap in the positive ($+$) and negative ($-$) helicity band being
\begin{equation}
\Delta^\pm_\mathbf{k} =\Delta^s \pm \Delta^t |\mathbf{l_k}|,
\end{equation}
respectively.
Thus, for sufficiently large $\Delta^t$, the negative-helicity gap $\Delta^-$ can change sign, i.e., line nodes generically appear on the negative-helicity Fermi surface~\footnote{If $\Delta^s$ and $\Delta^t$ have opposite sign the nodes appear on the positive-helicity Fermi surface.}.
This ensures the topological stability of flat bands of Majorana surface states within the projection of the bulk nodal lines onto the surface Brillouin zone~\cite{SBT12,STY11,BST11}.

\subsection{Setups}
\label{sec:setups}

Our goal is to find ways to confine and manipulate the Majorana modes. However, as these modes do not carry electric charge, one cannot hope to control them via an electric field. Instead, we will add an exchange field term to the Hamiltonian, which shifts the surface modes to non-zero energy by coupling to their spin polarization \cite{YSTY11,BTS13,MPSR13,STB13,WLLL13}. In the simplest setup, the exchange field is applied to a strip on the surface. In this scenario, we expect the zero-energy surface modes to be destroyed on the exchange-field strip, while they should persist at low energy in the field-free region.

\begin{figure}[!htbp]
\includegraphics[width=\columnwidth]{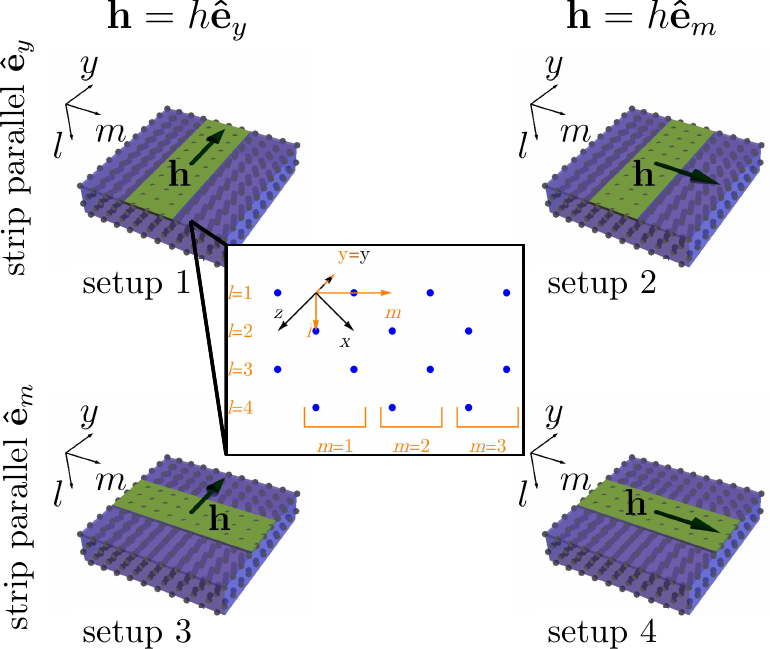}
\caption{Four different setups for the strip directions: The strip is oriented along the $y$ direction in setup 1 and setup 2 and along the $m$ direction in setup 3 and setup 4. The exchange field points in the $y$ direction in setups 1 and 3 and in the $m$ direction in setups 2 and 4.}
\label{fig:setups}
\end{figure}

To parametrize the slab with (101) surfaces, we keep the $y$ coordinate, which is parallel to the surface, while rotating the $x$ and $z$ directions to
\begin{align}
m&\equiv \left\lfloor\frac{x-z}{2}\right\rfloor, \\
l&\equiv x+z,
\end{align}
which are parallel and orthogonal to the slabs surface, respectively (see Fig. \ref{fig:setups}). We Fourier transform the BdG Hamiltonian in Eq.~\eqref{eq:H_BdG(k)} to real space and then consider a slab of thickness $L$ in the $l$ direction as well as length $M$ in the $m$ direction and $Y$ in the $y$ direction. The boundary conditions are open in the $l$ direction perpendicular to the surfaces and periodic in the $m$ and $y$ directions parallel to the surfaces. The exchange field can then be applied as a term $\mathbf{h}\cdot \boldsymbol{\hat{\sigma}}$ that acts on all surface-layer sites belonging to the exchange-field strip. This breaks translational invariance in the in-plane direction orthogonal to the strip  but respects it in the parallel direction. It is useful to leave the Hamiltonian in momentum space in the latter direction. A detailed derivation of the Hamiltonian is presented in Appendix~\ref{sec:Derivation of the mean-field Hamiltonian matrix}.  

For the concrete configuration of the strip and the exchange field, we consider the four main setups shown in Fig.~\ref{fig:setups}: both the strip and the field can point either in the $y$ direction or in the $m$ direction, which we will call setups~1,~2,~3, and~4, respectively. We do not consider out-of-plane fields, as they are more difficult to produce experimentally and do not yield any additional insight. Moreover, we ignore strips in any direction other than the two coordinate axes $m$ and $y$. We can, however, change whether the exchange field is applied to the $l=1$ surface or the $l=L$ surface, which will lead to different eigenstates and eigenvalues of the Hamiltonian because the $C_{4v}$ symmetry does not require the two surfaces to be equivalent.

\section{Spectrum and eigenstates for a strip}
\label{sec:Eigenvalues and eigenstates of the Hamiltonian}

In this section, we  examine the low-energy spectrum and the corresponding eigenstates of a system with an exchange field applied to a strip at the surface according to the four setups described in Sec.~\ref{sec:setups}. First, we construct a perturbative argument about the qualitative effect of the exchange field on the eigenvalues of the Hamiltonian. We then use exact diagonalization to confirm this hypothesis and reveal further details.

\subsection{First-order perturbation theory and spin polarization of the field-free system}
\label{sec:First-order perturbation theory and spin polarization of the field-free system}

For low field strengths $h=|\mathbf{h}|$, the exchange field term $\mathbf{h}\cdot \boldsymbol{\hat{\sigma}}$ can be considered as a perturbation to the field-free system. We label the states $|k_m,k_y,\nu\rangle$ of the field-free system according to their surface momentum $(k_m,k_y)$ and the index $\nu$, which enumerates the $4L$ states with the same surface momentum $(k_m,k_y)$, ordered by increasing modulus $|E|$ of the eigenenergy. Under the influence of an exchange field applied to the layer $l$, these states get shifted by an amount 
\begin{equation}
\Delta E_{|k_m,k_y,\nu\rangle} \propto \mathbf{h} \cdot \langle \mathbf{\hat{s}}_l
  \rangle_{|k_m,k_y,\nu\rangle},
\label{eq:ZeemanE}
\end{equation}
up to first order of perturbation theory. In this equation, the expectation value $\left\langle \boldsymbol{\hat{\sigma}}_{l}\right\rangle_{|k_m,k_y,\nu\rangle}
$ of the spin polarization is defined as
\begin{equation}
\left\langle \mathbf{\hat{s}}_l \right\rangle_{|k_m,k_y,\nu\rangle}=\langle k_m,k_y,\nu|P_{l,l}\otimes \begin{pmatrix}
\boldsymbol{\hat{\sigma}}&0\\0&-\boldsymbol{\hat{\sigma}}^\top
\end{pmatrix}|k_m,k_y,\nu\rangle,
\end{equation}
where $P_{l,l}$ is an $L{\times}L$ matrix with entry $1$ in the $(l,l)$ component and zero entries otherwise. Thus, to the first order of perturbation theory, the energy corrections due to the applied exchange field are proportional to the zero-field spin polarization.
 
\begin{figure*}[!htbp]
\includegraphics[width=\textwidth]{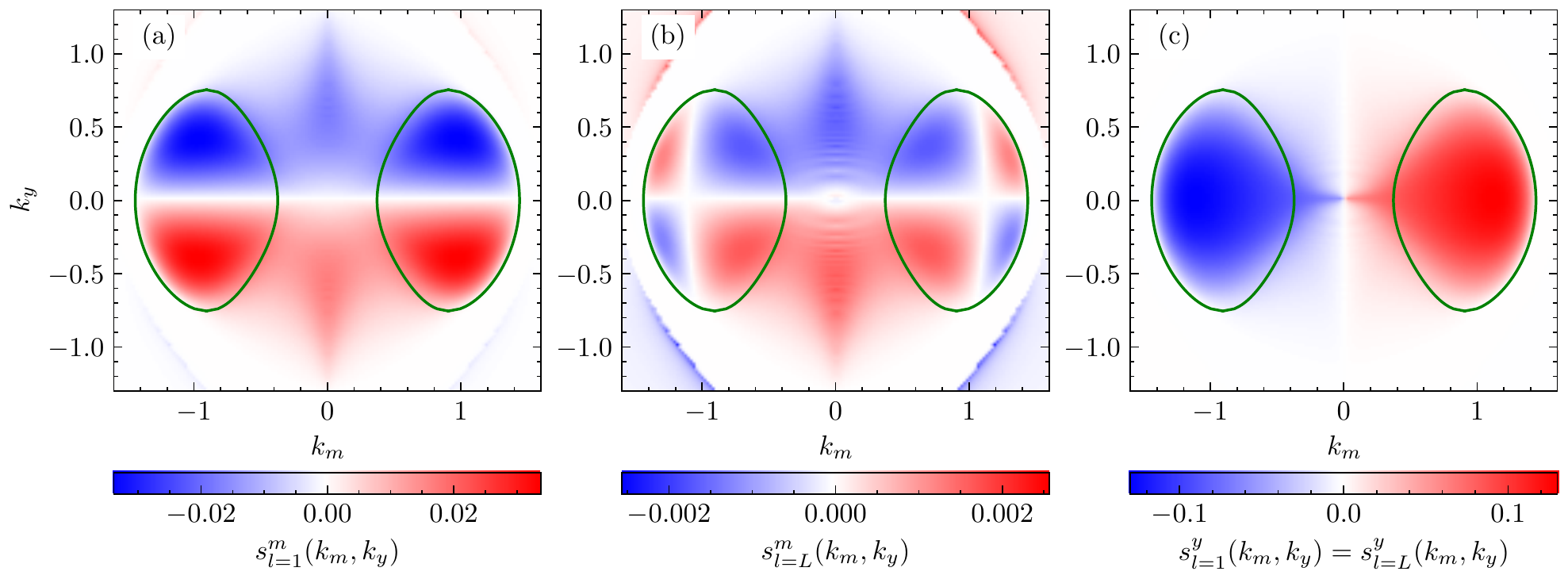}
{\phantomsubcaption \label{subfig:spinm1}
\phantomsubcaption \label{subfig:spinmL}
\phantomsubcaption \label{subfig:spiny1}}
\caption{Spin polarization of the field-free system. \subref{subfig:spinm1} $m$ component of the spin at the $l=1$ surface and \subref{subfig:spinmL} at the $l=L$ surface, as a function of the surface momenta $k_m$ and $k_y$. \subref{subfig:spiny1} $y$ component of the spin, which is equal at the $l=1$ and the $l=L$ surface. The green lines are the projections of the bulk nodal lines onto the surface Brillouin zone. As the maximal spin polarization differs significantly between the three cases, different color scales are used.}
\label{fig:spin}
\end{figure*}

The momentum-dependent spin polarization of the field-free system can be calculated by transforming the BdG Hamiltonian in Eq.~\eqref{eq:H_BdG(k)} to real space in the $l$ direction perpendicular to the slab and using open boundary conditions (see Appendix~\ref{sec:Derivation of the mean-field Hamiltonian matrix}). Figure~\ref{fig:spin} shows the result for the $m$ and $y$ components of the spin at the $l=1$ surface and for the $m$ component at the $l=L$ surface of a slab with thickness $L=200$, hopping amplitude $t=1$, spin-orbit coupling $\lambda=-1.5$, chemical potential $\mu=-3$, and gaps $\Delta^s=0.3426$ and $\Delta^t=0.5$. These parameters are also used for all further calculations. Our qualitative results do not depend on the specific values of these parameters. The $y$ components $s^y_{l}$ of the spins are the same for $l=1$ and $l=L$. Hence, for setups 1 and 3, we do not expect different shifts in energy for the two surfaces. In general, we expect the energy shift to be linear in the field strength $h$ for all those momenta in the strip direction (i.e., $k_y$ in setups 1 and 2 and $k_m$ in setups 3 and 4), for which the corresponding spin polarization is nonzero. Thus, the momentum $k_y=0$ in setup 2 is an exception, for which this argument does not hold, and which will therefore not show a linear field dependence of the energy shift, because $s^{(m)}_{l=1}$ and $s^{(m)}_{l=L}$ vanish by symmetry.

Moreover, the $m$ component of the spin at the $l=L$ surface is much smaller than for the $l=1$ surface and there are additional sign changes for $k_m$ close to $\pm 1$. The origin of this is an accidental near cancellation of spin polarizations in the \textit{x} and \textit{z} directions. The weak spin polarization and corresponding small energy shifts have a profound effect on the surface states, as we will see below.

\subsection{Classification of surface states}
\label{sec:Classification of surface states}

All calculations in this section are performed for the exchange-field strip covering half the surface of width $M=100$ in setups 1 and 2 and $Y=100$ in setups 3 and 4 , i.e., the exchange-field strip has a width of $\Delta m=M/2=50$ in setups 1 and 2 and of $\Delta y=Y/2=50$ in setups 3 and 4. The strip is centered in the middle of the slab, i.e., at $m=50$ or $y=50$.

\begin{figure}[!htbp]
\includegraphics[width=0.48\textwidth]{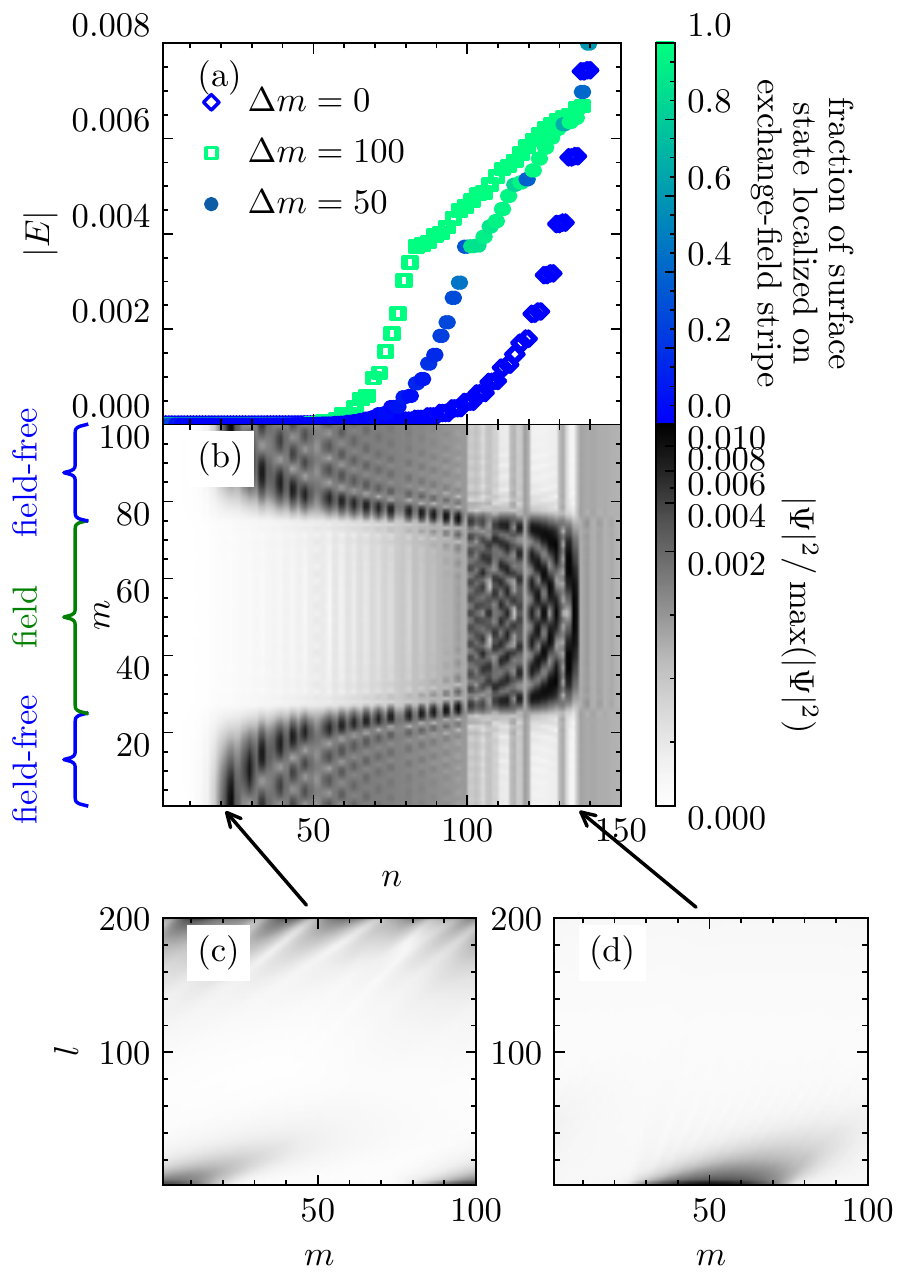}
{\phantomsubcaption \label{subfig:box_antibox_a}
\phantomsubcaption \label{subfig:box_antibox_b}
\phantomsubcaption \label{subfig:box_antibox_c}
\phantomsubcaption \label{subfig:box_antibox_d}
}
\caption{Surface states and their energies arising for setup 1, which exhibits nonzero spin polarization in the field-free system. \subref{subfig:box_antibox_a} Energies at $k_y=0$ of a slab with an exchange field $\mathbf{h}=0.025\, \mathbf{\hat{e}}_y$ applied to a strip along the $y$ direction at the $l=1$ surface, arranged in increasing order and colored according to the fraction of the squared modulus of the wave function that is localized in the exchange-field part of the surface. For reference, the field-free case (diamonds) and the fully covered surface (squares) are plotted as well. \subref{subfig:box_antibox_b} Squared modulus of the corresponding wave function at $l=1$. \subref{subfig:box_antibox_c} Probability density of the first box state and \subref{subfig:box_antibox_d} of the last anti-box state over the whole thickness $l\in\lbrace 1, \hdots, L\rbrace$ of the slab.}
\label{fig:box_antibox}
\end{figure}

Figure~\ref{fig:box_antibox} shows the energies and probability densities of surface states for an example of the first situation explained in Sec.~\ref{sec:First-order perturbation theory and spin polarization of the field-free system}, i.e., one with a nonzero spin polarization in the exchange-field direction. The calculations for this example are performed for the surface momentum $k_y=0$ and a field $\mathbf{h}=0.025\, \mathbf{\hat{e}}_y$ applied according to setup 1 at the $l=1$ surface, i.e., with both the field and the strip oriented along the $y$ direction. As the Hamiltonian is a $4LM{\times}4LM$ matrix [see Eq.~\eqref{eq:H_matrix_set12} in Appendix~\ref{sec:Derivation of the mean-field Hamiltonian matrix}], each vector representing an eigenstate $\Psi$  can be divided into $LM$ tuples of length four. Thus, every site $(m,l)$ corresponds to a quadruple
\begin{align}
\Psi_{n}(m,l) &= (\Psi^{(p,\uparrow)}_{n}(m,l),\Psi^{(p,\downarrow)}_{n}(m,l), \nonumber \\
&\qquad \Psi^{(h,\uparrow)}_{n}(m,l),\Psi^{(h,\downarrow)}_{n}(m,l)),
\end{align}
which represents the particle-spin-up, particle-spin-down, hole-spin-up, and hole-spin-down amplitudes of the state at the site $(m,l)$.

The states in Figs.~\ref{subfig:box_antibox_a} and (b) are ordered according to increasing absolute value of the corresponding energy $|E|$ and enumerated by an index $n$. Figure~\ref{subfig:box_antibox_b} shows the $l=1$ part of the squared modulus
\begin{equation}
|\Psi_{n}(m,l=1)|^2 \equiv \sum_{i=(p,\uparrow),(p,\downarrow),(h,\uparrow),(h,\downarrow)}|\Psi^{i}_{n}(m,l=1)|^2
\end{equation} 
of the wave function as a function of $m$ (vertical axis) from $n=1$ to $150$ (horizontal axis). In Fig.~\ref{subfig:box_antibox_a}, the corresponding eigenvalues are plotted and colored according to the fraction $\sum_{m=26}^{75}|\Psi_{n}(m,l=1)|^2/\sum_{m=1}^{100}|\Psi_{n}(m,l=1)|^2$ of the surface part of the state that is localized on the exchange-field strip. The spectra plotted using empty diamonds and empty squares in Fig.~\ref{subfig:box_antibox_a} refer to the cases with zero applied field and with the field applied to the whole surface, respectively. So, compared to the full-field case, only approximately half as many states at the $l=1$ surface get shifted away from zero energy. All surface states decay rapidly into the bulk, as can be seen exemplarily in Figs.~\ref{subfig:box_antibox_c} and (d). The lowest eigenvalues $|E_n|$ in Fig.~\ref{subfig:box_antibox_a} correspond to states localized at the opposite, field-free $l=L$ surface and will be ignored in further discussion. Starting at $n=21$, the states are localized almost entirely on the field-free strip. They resemble the states of a quantum mechanical particle in a box potential in that they have an increasing number of nodes with increasing energy and decay rapidly into the exchange-field strip, i.e., the walls of the box. We will call these states \emph{box states} from now on. As an example, a state with zero nodes that is localized almost entirely on the field-free strip is depicted in Fig.~\ref{subfig:box_antibox_c}. At higher energies, a non-negligible part of the states starts to be localized at the boundaries between the two kinds of strips and on the exchange-field strip, until finally, there is a sharp transition at $n\approx 100$. Beyond this point, the states are localized mostly on the exchange-field strip and, similar to the low-energy states introduced above, the number of nodes depends on $n$. However, in this case, the number of nodes \emph{decreases} with increasing energy. Therefore, we are going to call these states \emph{anti-box states} from now on. The last of these anti-box states, which has zero nodes, is shown in Fig.~\ref{subfig:box_antibox_d}.

\begin{figure}[!htbp]
\includegraphics[width=0.48\textwidth]{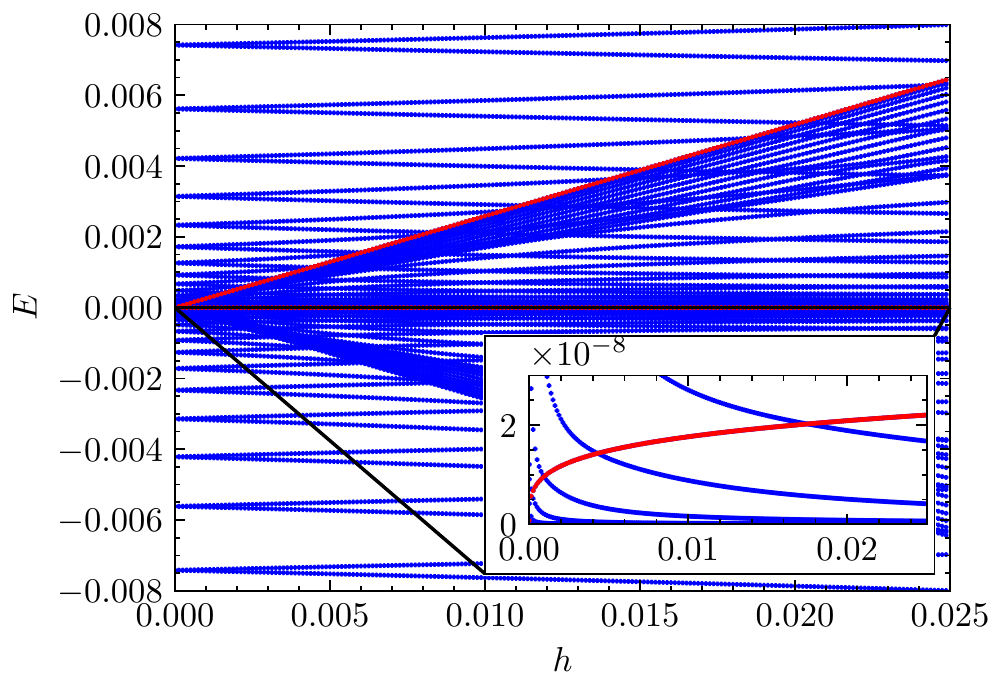}
\caption{Energies corresponding to the surface states at $k_y=0$ for an exchange field according to setup 1 applied to the $l=1$ surface, for varying exchange field strength $h$. The field dependence of the eigenvalues corresponding to the highest anti-box state and the lowest box state are indicated in red. Inset: Zoom-in on the energy of the lowest box state.}
\label{fig:E(h)lin}
\end{figure}

According to the perturbative arguments in Sec.~\ref{sec:First-order perturbation theory and spin polarization of the field-free system}, the eigenenergies should be linear in the field strength. Figure \ref{fig:E(h)lin} shows that for $k_y=0$, the field dependence of the eigenenergies is indeed linear for the anti-box states, while it stays very close to zero for the box states. The shift of the eigenenergy for the lowest box state is several orders of magnitude smaller than for the anti-box states. In contrast to the anti-box states, the field dependence of the box states is not linear. Instead, the initial increase is characterized by an exponent that is smaller than unity and the curve flattens for stronger fields. This can be attributed to the fact that even though the box state is mostly localized on the field-free strip, it has a small nonzero weight on the exchange-field strip. This part of the state is strongly affected by the exchange field and leads to a small energy shift. However, the weight of the box state on the exchange-field strip decreases with increasing field strength, which leads to the nonlinear behavior. As examples for  both the box and the anti-box states, the field dependence of the eigenenergies of the states with zero nodes on the exchange-field strip and on the field-free strip are indicated in red in Fig.~\ref{fig:E(h)lin}.

\begin{figure*}[!htbp]
\includegraphics[width=0.8\textwidth]{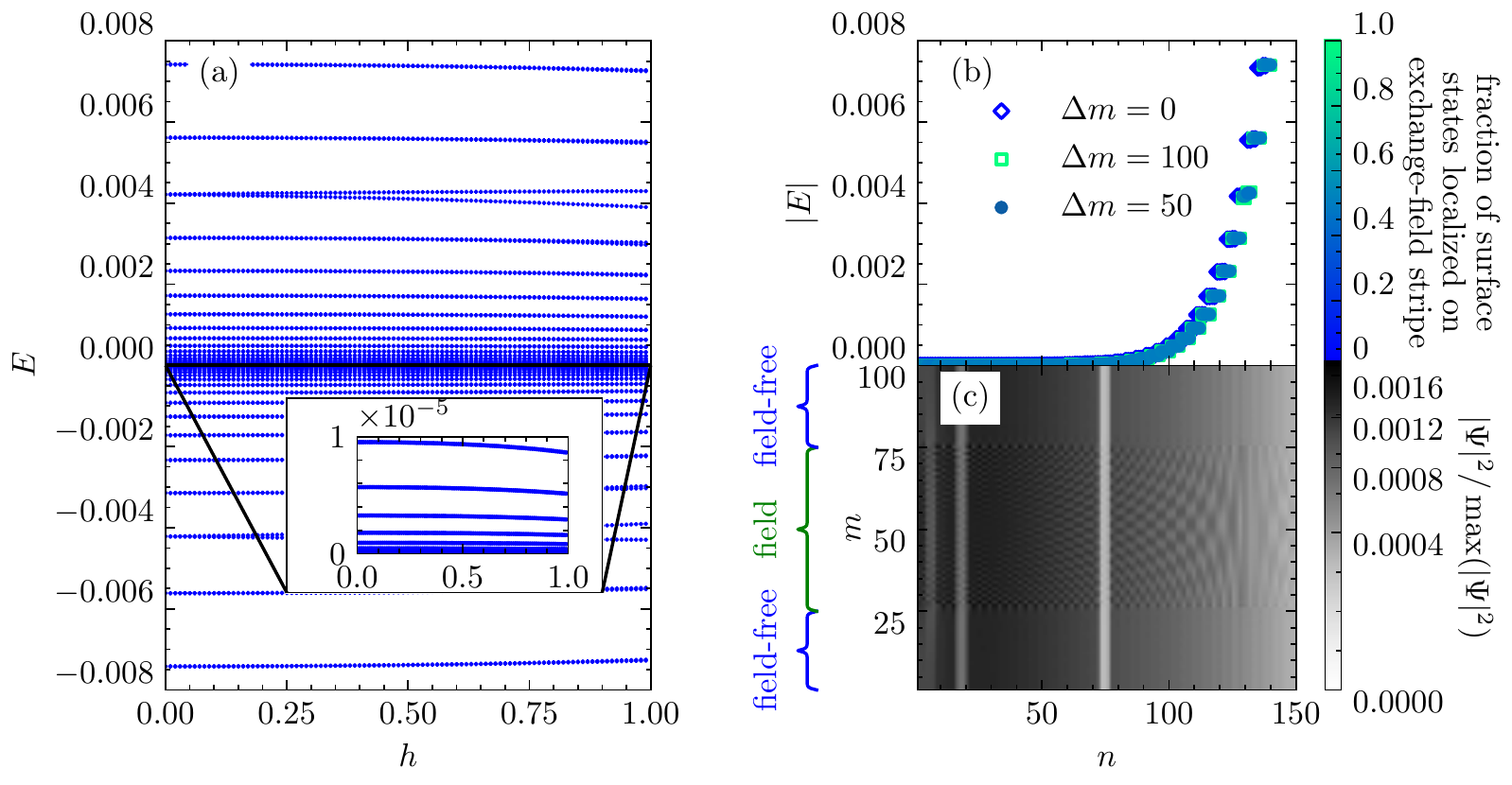}
{\phantomsubcaption \label{subfig:set2_a}
\phantomsubcaption \label{subfig:set2_b}
\phantomsubcaption \label{subfig:set2_c}}
\caption{Energies and surface states at $k_y=0$ for an exchange field according to setup 2 applied to the $l=1$ surface. \subref{subfig:set2_a} Spectrum of the Hamiltonian for varying exchange field strength $h\in[ 0,1]$. \subref{subfig:set2_b} Eigenenergies for an exchange field $\mathbf{h}=0.25\, \mathbf{\hat{e}}_m$, colored according to the fraction of the squared modulus of the wave function which is localized on the field strip and arranged in increasing order.  For reference, the field-free case (diamonds) and the fully covered surface (squares) are plotted as well. \subref{subfig:set2_c} Squared modulus of the wave function of the corresponding eigenstates in the $l=1$ layer.}
\label{fig:set2}
\end{figure*}

For all other setups, results of the analogous calculations are qualitatively similar to setup 1 in that they exhibit low-energy box states on the field-free strip and anti-box states with linearly field-dependent energy on the exchange-field strip, with two notable exceptions. One of these is a field applied according to setup 2 for states at $k_y=0$. Results for this situation are depicted in Fig.~\ref{fig:set2}. As shown in Fig.~\ref{subfig:set2_a}, there are no anti-box states with linearly field-dependent eigenvalues. All flat-band states remain close to zero energy, the lowest order of field dependence is quadratic, and there is only a weak shift in energy even at high field strength, which can be seen in Fig.~\ref{subfig:set2_b} for $h=0.25$. Moreover, the states are not clearly localized on either one of the two strips [see Fig.~\ref{subfig:set2_c}]. This deviation from the previously described behavior results from the fact that the relevant spin polarization for this setup is zero for all states at $k_y=0$. Thus the correction to the eigenenergies of first order in the field vanishes and the remaining field dependence is quadratic.

\begin{figure}[!htbp]
\includegraphics[width=0.48\textwidth]{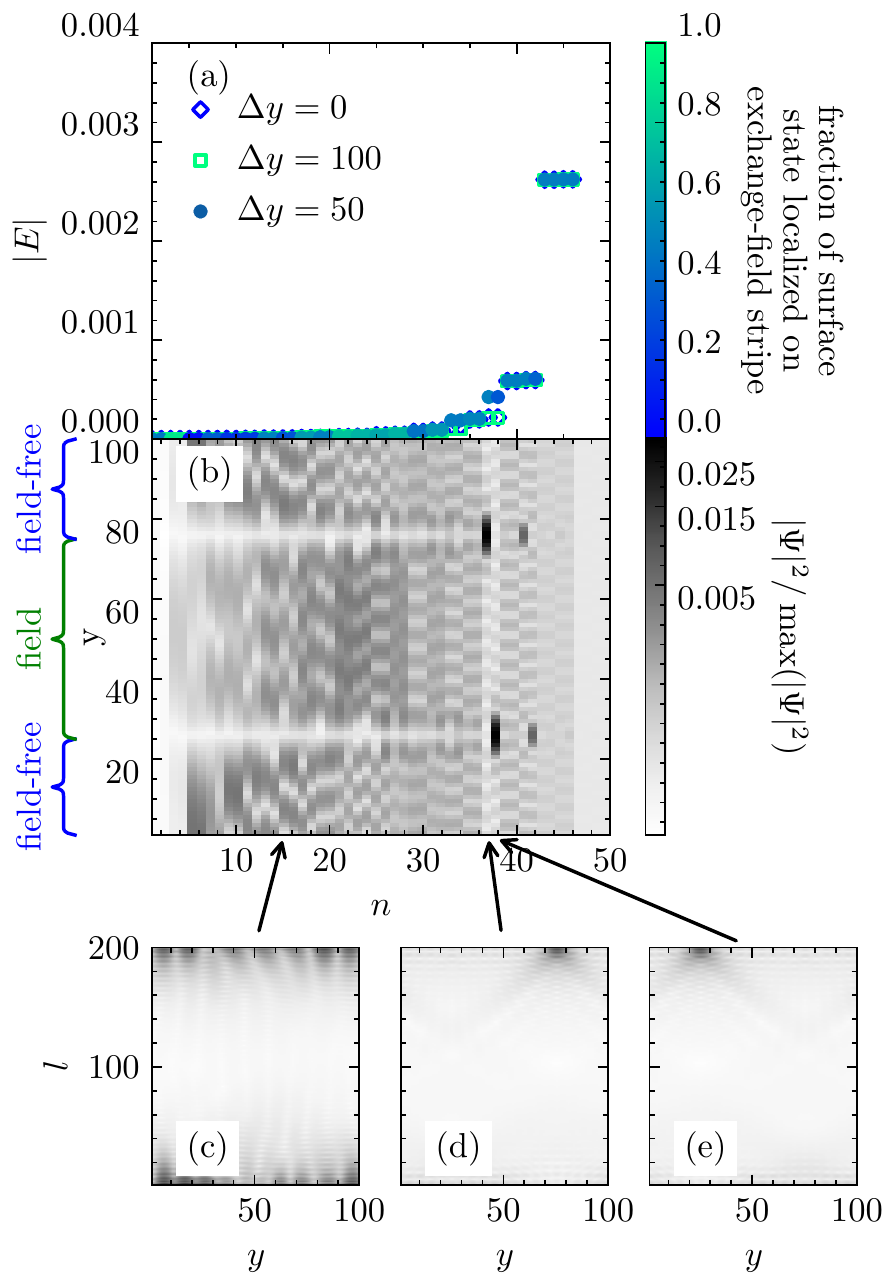}
{\phantomsubcaption \label{subfig:set4d_a}
\phantomsubcaption \label{subfig:set4d_b}
\phantomsubcaption \label{subfig:set4d_c}
\phantomsubcaption \label{subfig:set4d_d}
\phantomsubcaption \label{subfig:set4d_e}}
\caption{Surface states and energies arising from a field according to setup 4 at the $l=L$ surface of the slab. \subref{subfig:set4d_a} Energies at $k_y=1$, colored according to the fraction of the squared modulus of the wave function which is localized on the field strip of the surface and arranged in increasing order. For reference, the field-free case (diamonds) and the fully covered surface (squares) are plotted as well. \subref{subfig:set4d_b} Squared modulus of the corresponding wave function at $l=L$. \subref{subfig:set4d_c} Example of a surface state that is localized on both of the strips. \subref{subfig:set4d_d}, \subref{subfig:set4d_e} The two states that are localized on either of the strip's boundaries.}
\label{fig:set4d}
\end{figure}

\begin{figure}[!htbp]
\includegraphics[width=0.48\textwidth]{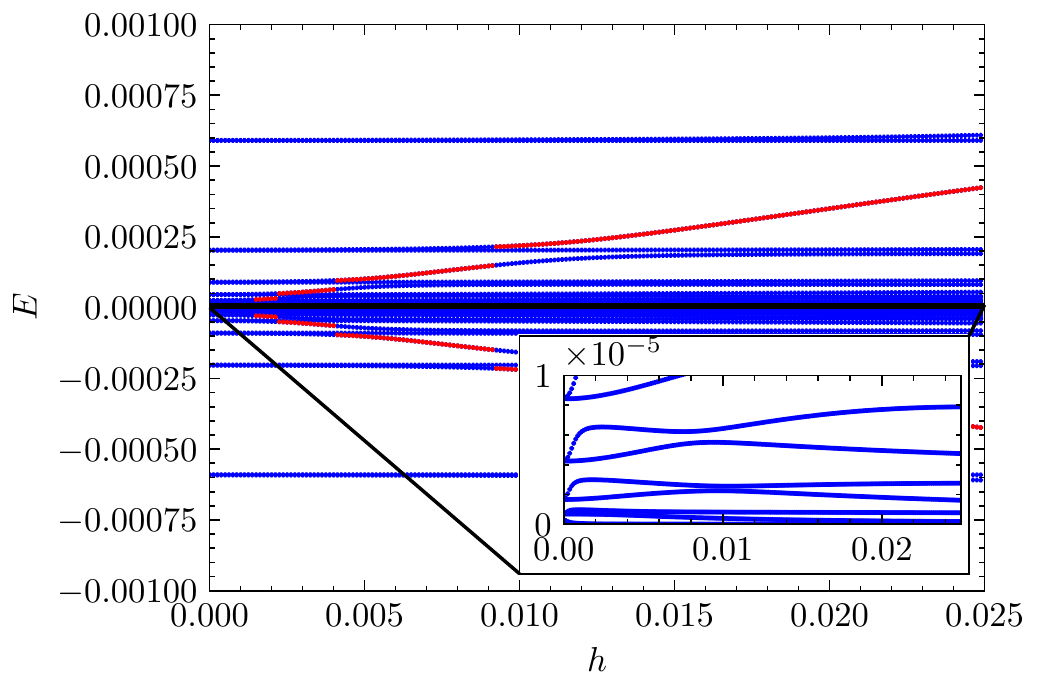}
\caption{Energies corresponding to the surface states of a system at $k_m=1$ for which an exchange field according to setup 4 is applied to the $l=L$ surface for varying exchange-field strength $h$. The field dependence of the eigenvalues corresponding to the states localized on the boundaries between the two strips are indicated in red. Inset: Zoom-in on the low-energy part of the spectrum.}
\label{fig:set4dfield}
\end{figure}

The other exception from the box/anti-box phenomenology occurs if both strip and field are oriented along the $m$ direction at the $l=L$ surface, which is shown in Fig.~\ref{fig:set4d} for $k_m=1$. In this case, most states remain at low energy and $|\Psi_{n}(m,l=1)|^2$ oscillates strongly at the entire surface of the slab. Thus, they are not localized on either one of the strips [see Figs.~\ref{subfig:set4d_a} and (b)]. An example of such a state is shown in Fig.~\ref{subfig:set4d_c}. The only two surface states that do not obey this pattern are two states localized on either of the boundaries between the two kinds of strips, shown in Figs.~\ref{subfig:set4d_d} and (e). The eigenenergy of one of these states has a linear field dependence with a positive slope, while the other has a negative slope with the same modulus (see Fig.~\ref{fig:set4dfield}). These observations can be explained as follows: For every state originating from a Majorana surface mode at $(k_m,k_y)$, the first-order perturbation theory has to start from a linear combination of the two degenerate states at $(\pm k_m,k_y)$. As shown by Fig.~\ref{subfig:spinmL}, for $k_m=1$ these states have a small zero-field $m$-spin polarization, which, however, strongly oscillates between positive and negative values along the real-space $m$ axis. A derivation of this fact can be found in Appendix~\ref{sec:Oscillating m-spin polarization for setup 4 on the l=L surface}. If a field is applied, none of these oscillations lead to an actual shift of the eigenenergy of a state because the positive and negative spins cancel out. Therefore, almost all states remain at zero energy. Only in the case where a peak of the $m$-spin polarization is on one side of the boundary between the field-free and the exchange-field strip and the corresponding dip is on the other, does the energy get shifted upward or downward linearly. Thus, the two states that are shifted linearly are strongly localized at the boundary between the two kinds of strips.

To conclude this section, we emphasize that generically the strips exhibit a dichotomy of box states with weak field dependence and anti-box states with linear field dependence. Exceptions occur if the relevant spin polarization for the field-free surface either vanishes exactly because of symmetry or is accidentally small.

\section{Dispersion in the strip direction}
\label{sec:Dispersion in strip direction}

\begin{figure*}[!htbp]
\setlength{\tabcolsep}{2pt}
\begin{tabular}{l|l|l}
\toprule
	&$\mathbf{h}=h\mathbf{\hat{e}}_y$&$\mathbf{h}=h\mathbf{\hat{e}}_m$\\ \midrule
\rotatebox{90}{strips $\parallel \mathbf{\hat{e}}_y$}&\includegraphics[width=0.46\textwidth]{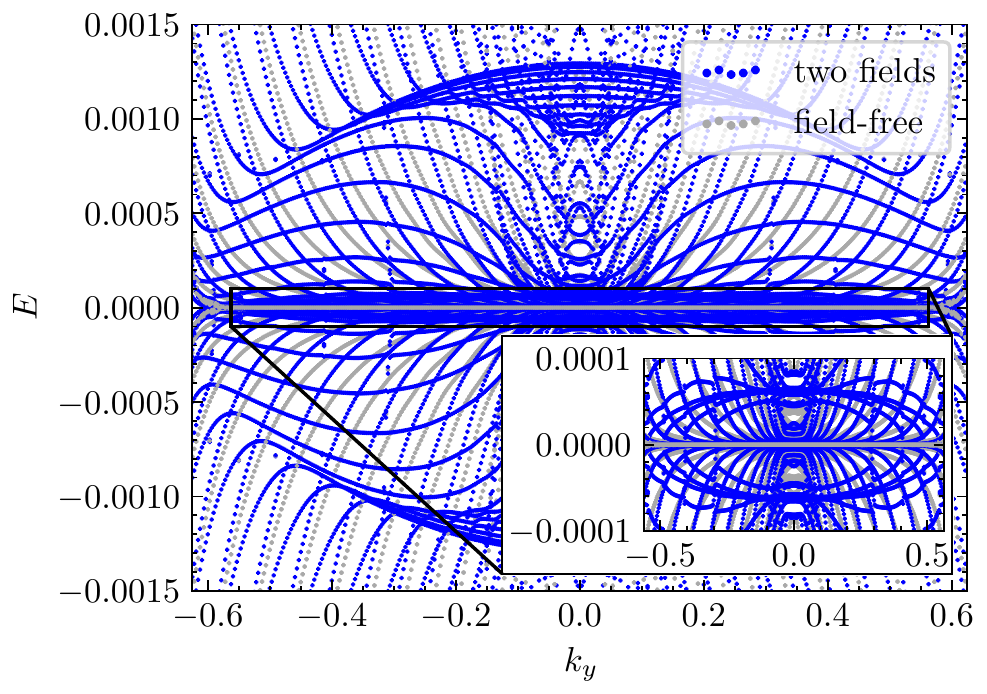}&\includegraphics[width=0.46\textwidth]{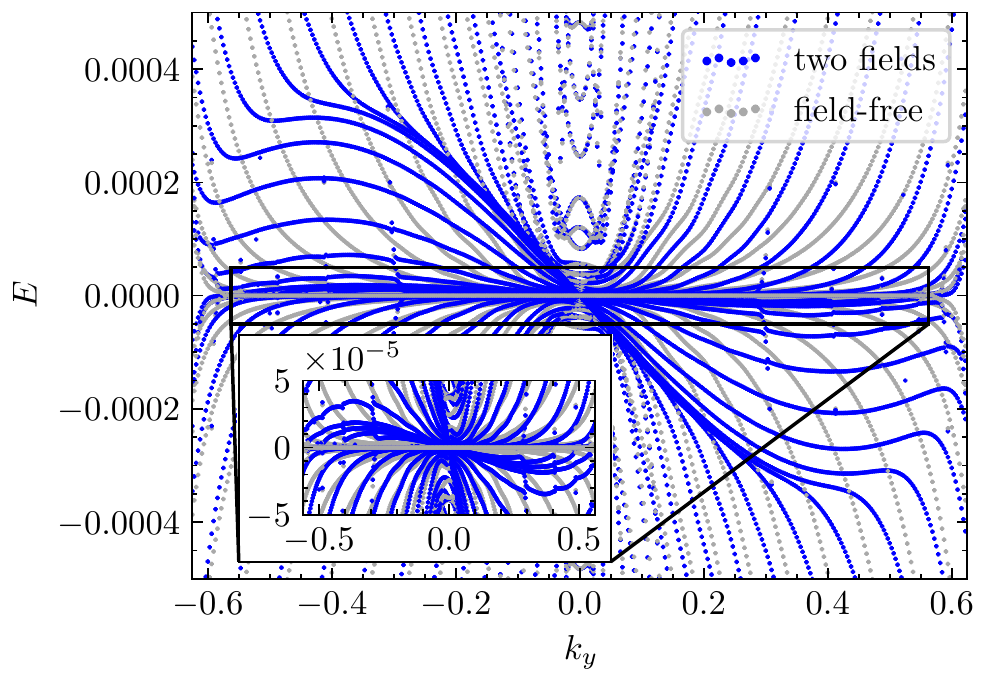}\\
	& (a) setup~1 & (b) setup~2\vspace{2pt}\\ \hline
	\rotatebox{90}{strips $\parallel \mathbf{\hat{e}}_m$} &\includegraphics[width=0.46\textwidth]{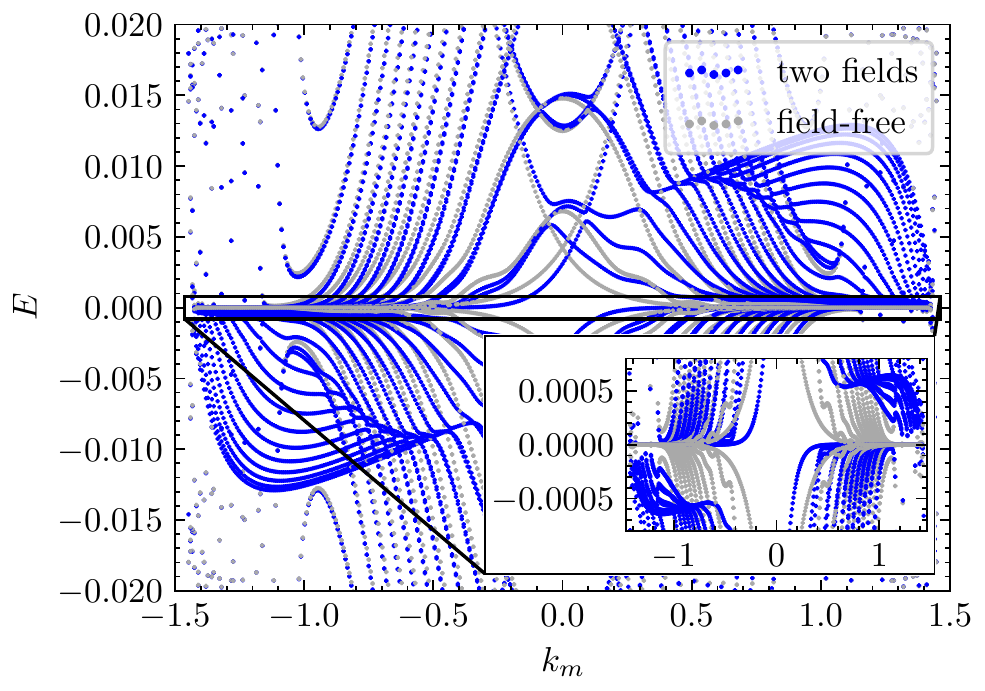}& \includegraphics[width=0.46\textwidth]{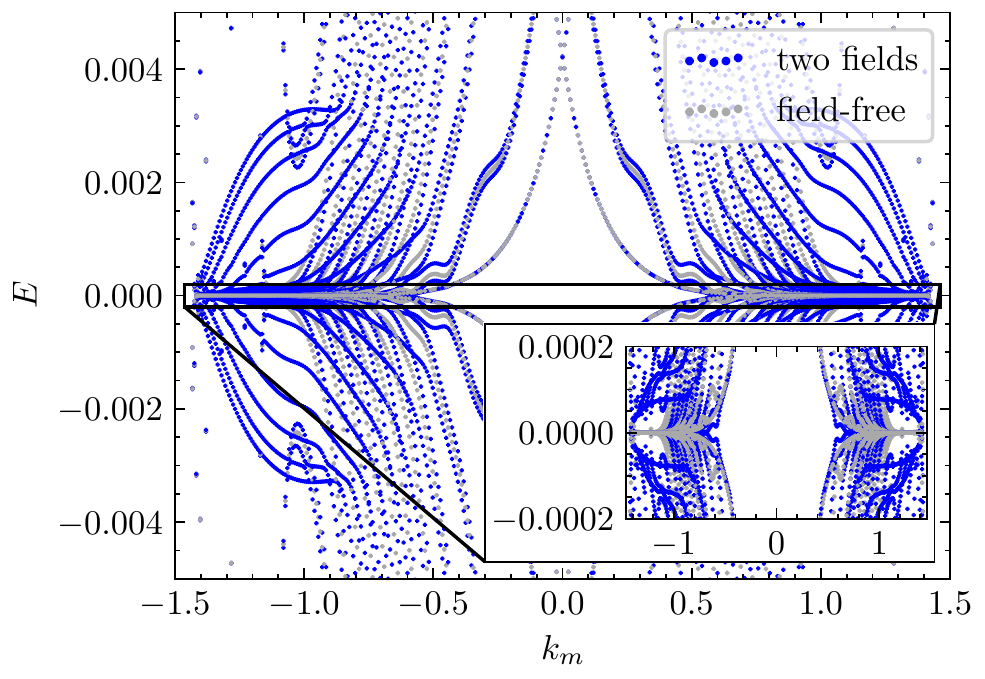}\\
& (c) setup~3 &  (d) setup~4\\
	\bottomrule
\end{tabular}
{\phantomsubcaption \label{subfig:twofields_set1}
\phantomsubcaption \label{subfig:twofields_set2}
\phantomsubcaption \label{subfig:twofields_set3}
\phantomsubcaption \label{subfig:twofields_set4}
}
\caption{Dispersion for the specified setups, where a strong field is applied to one strip at the $l=1$ surface of the slab to restrict the flat-band states to the other strip. A weak field is applied to the latter strip, in order to make the previously flat bands weakly dispersive. Both fields point in the same direction, i.e., the $y$ direction for panels \subref{subfig:twofields_set1} and \subref{subfig:twofields_set3} and the $m$ direction for panels \subref{subfig:twofields_set2} and \subref{subfig:twofields_set4}, while the strip is oriented along the $y$ direction for panels \subref{subfig:twofields_set1} and \subref{subfig:twofields_set2} and the $m$ direction for panels \subref{subfig:twofields_set3} and \subref{subfig:twofields_set4}. The dispersion is plotted in blue. For reference, the dispersion of a completely field-free system is given in light gray in the background. The insets show zoom-ins on the low-energy parts of the dispersions.}
\label{fig:twofields_setups}
\end{figure*}

In this section, we propose a method to move localized wave packets of Majorana zero modes. As described in Sec.~\ref{sec:intro}, one can form wave packets from the zero-energy Majorana modes within the projection of the bulk nodal lines. As shown in Sec.~\ref{sec:Classification of surface states}, one can localize modes at a strip on the surface by applying an exchange field everywhere else, which will shift states with a significant weight outside of the field free strip (i.e., anti-box states) to high energies. The box states with most of their weight on the field-free strip remain at much lower energy. They are still \emph{approximately} degenerate in the conserved momentum component parallel to the strip and so one can choose an appropriate linear combination of the box states at different momenta to build a localized wave packet. The simplest idea for moving the wave packets is to apply a small field to the previously field-free strip in order to create a weak linear dispersion along the strip \cite{BTS13,STB13}. A wave packet would move without broadening if it were a superposition of components with the same velocity $\partial E/\partial k_\mathrm{strip}$ along the strip, where $k_\mathrm{strip}$ is the momentum parallel to the strip.

One can make predictions about the resulting dispersion based on the spin polarization of the field-free system. For every momentum $k_\text{strip}$, the first-order perturbation theory described in Sec.~\ref{sec:First-order perturbation theory and spin polarization of the field-free system} can be applied, i.e., the exchange field couples to the spin polarization of the surface states of the \emph{field-free} system. For every value of $k_\text{strip}$, there are fewer than $M$ (setup 1 or 2) or fewer than $Y$ (setup 3 or 4) surface states corresponding to different values of the momentum $k_{\perp}$ orthogonal to the strip direction in the field-free system. If the field is switched on, approximately half of these states are shifted away from zero energy by an amount $\Delta E$ which, to first order, is proportional to the spin polarization. The other half correspond to the other surface of the slab. Thus, the shape of the resulting dispersion $E(k_\text{strip})$ can be predicted from the projection of the zero-field spin polarization along the direction orthogonal to the strip.

In Fig.~\ref{fig:twofields_setups}, the dispersion $E(k_\text{strip})$ is shown for all four setups. The large field was chosen as $h_\text{large}=0.005$ for setups 1 and 2 and $h_\text{large}=0.05$ for setups 3 and 4, while the small field was $h_\text{small}=0.00025$ for setups 1 and 2 and $h_\text{small}=0.0025$ for setups 3 and 4. The dispersion in Fig.~\ref{subfig:twofields_set1}, which belongs to setup 1, is inappropriate for our goals, as the bands are merely shifted away from zero energy instead of being tilted to form a linear dispersion. On the other hand, setup 2 in Fig.~\ref{subfig:twofields_set2} displays a linear dispersion over a wide range of momenta $k_y$, which should allow one to move a wave packet in the $y$ direction. Since the dispersion is not perfectly linear, the wave packet will broaden with time.

To move a wave packet in the $m$ direction, it would be necessary to construct a linear dispersion on the small-field strip of either setup 3 or setup 4. Similar to setup 1, setup 4 does not lead to a linear dispersion, but rather shifts the bands away from zero energy, as can be seen in Fig.~\ref{subfig:twofields_set4}. For setup 3, parts of the dispersion are linear, as shown in Fig.~\ref{subfig:twofields_set3}. However, we see that there are no linearly dispersing low-energy states at small momentum $k_m$. The reason for this is clear from Fig.~\ref{fig:spin}: There are no flat-band states at small $k_m$ for the field-free surface \footnote{Even if the system parameters are such that the two projections of the bulk nodal lines touch or overlap, there will be at most two points of zero-energy states at $k_m=0$, because the momentum dependent winding number that protects the zero-energy Majorana modes within the projection of the bulk nodal lines cancels if two of those regions overlap. Thus, only the boundaries of two of these regions will touch at $k_m=0$.} It is nevertheless possible to construct a wave packet out of states with approximately the same velocity. However, the high degree of anisotropy between strips in the two orthogonal directions on the surface is likely detrimental to constructing more complicated structures.

A crucial insight is that the general shape of the dispersions of the introduced two-field setup can be predicted from the spin polarization of the field-free system. In first-order perturbation theory, the dispersion in the strip direction is determined by the projections of the spin polarizations along the axis orthogonal to the strip. The spin polarization of the field-free system is thus a straightforward tool to predict which point groups other than $C_{4v}$ are promising candidates to achieve weak linear dispersions on strips in two linearly independent directions. Figure 4 of Ref.~\cite{BST15} shows that for the (111) surface of a NCS with point group $O$, the spin polarization is parallel to the surface and rotates by $2\pi$ when the momentum parallel to the surface is rotated by $2\pi$. This system is thus promising for strips in arbitrary directions, which we leave for future research.

\section{Summary and conclusions}
\label{sec:summary}

Motivated by the goal of manipulating localized Majorana modes at the surface of a NCS, we have analyzed the consequences of the application of an exchange field to part of a surface with Majorana flat bands. As a model system, we have used a slab with (101) surfaces of a superconductor with point group $C_{4v}$ with an exchange field applied to a strip on the surface.

We have seen both from first-order perturbation theory and exact diagonalization that in cases where the spin polarization of the zero-energy surface states in the field-free system is not zero or very small for a certain momentum in the strip direction, the eigenstates at low energies are localized on the field-free strip. They have an increasing number of nodes with weakly increasing energy, and thus resemble the states of a particle in a box, which is why we have called them box states. On the other hand, there are states which are shifted away from zero energy by an amount $\Delta E$ that is proportional to the exchange-field strength $h$. These states are localized on the exchange-field strip. At fixed field strength, they have a decreasing number of nodes with increasing energy and we have called them anti-box states. It is thus possible to achieve the first prerequisite for manipulating the surface modes: to constrain them to predefined regions. This picture breaks down if the spin polarization of the field-free surface states vanishes by symmetry or is accidentally small. In this case, first-order perturbation theory is no longer valid and we do not find well-defined box and anti-box states.

We have also considered a small exchange field on the previously field-free strip with the goal to introduce a linear dispersion to the almost flat bands of Majorana modes. We have found that it is possible to obtain an approximately linear dispersion for a range of momenta for strips in both the $y$ and the $m$ directions. Hence, by switching the weak field on and off, one can, in principle, also achieve the second prerequisite for Majorana manipulation: to move wave packets in a controlled manner. The deviation from perfect linearity will lead to broadening of wave packets with time. By making the support of the wave packets narrower in momentum space, they become broader in real space but the velocities become more uniform so that the additional time-dependent broadening is reduced. The necessary optimizing of wave packets and the dynamics resulting from switching the weak field on and off are interesting topics for future research. In general, the shape of the dispersion on the weak-field strip can be predicted from the spin polarization of the field-free system. Thus, good candidates for model systems and setups that allow for a linear dispersion in two independent surface directions can be identified based on the spin polarization.

\vspace*{3ex}
\acknowledgments

We thank P. M. R. Brydon and J. E. R\"uckert for useful discussions. Financial support by the Deut\-sche Forschungsgemeinschaft via the Collaborative Research Center SFB 1143, project A04, and the Cluster of Excellence on Complexity and Topology in Quantum Matter ct.qmat (EXC 2147) is gratefully acknowledged.

\appendix

\section{Derivation of the mean-field Hamiltonian matrix} \label{sec:Derivation of the mean-field Hamiltonian matrix}

In this appendix, we summarize the derivation of the Hamiltonian matrix that we use to obtain the eigenstates and eigenenergies in Secs.~\ref{sec:Classification of surface states} and \ref{sec:Dispersion in strip direction}. We first transform the Hamiltonian in Eq.~\eqref{eq:H_BdG(k)} from the momenta $(k_x,k_y,k_z)$ to $(k_m,k_y,k_l)$ via the relation
\begin{equation}
\begin{pmatrix}
k_x\\k_y\\k_z
\end{pmatrix}
=
\begin{pmatrix}
\frac{1}{\sqrt{2}}  & 0 & \frac{1}{\sqrt{2}}\\
0                   & 1 & 0                 \\
-\frac{1}{\sqrt{2}} & 0 & \frac{1}{\sqrt{2}}
\end{pmatrix}
\begin{pmatrix}
k_m\\k_y\\k_l
\end{pmatrix}.
\end{equation}
Then, we perform a Fourier transformation
\begin{equation}
c_{(k_m,k_y,k_l),\sigma}=\frac{1}{\sqrt{L}}\sum_{l\in \lbrace 1, \hdots, L \rbrace}e^{-i k_l l/\sqrt{2}}c_{(k_m,k_y,l),\sigma}
\end{equation}
in the $l$ direction with open boundary conditions, which leads to 
\begin{widetext}
\begin{align}
H &= \frac{1}{2L} \sum_{k_m, k_y, l, l^\prime} \Psi_{(k_m,k_y,l)}^\dagger e^{i k_l (l-l^\prime)/\sqrt{2}}
\begin{pmatrix}
 -\mu -2 t \cos(k_y) & \lambda  \sin(k_y) & -\Delta^t \sin(k_y) & \Delta^s\\ 
 \lambda  \sin(k_y) & -\mu -2 t \cos(k_y) & -\Delta^s & \Delta^t \sin(k_y) \\
 -\Delta^t \sin(k_y) & \Delta^s & \mu +2 t \cos(k_y) & -\lambda  \sin(k_y) \\
 -\Delta^s & \Delta^t \sin (k_y) & -\lambda  \sin(k_y) & \mu + 2 t \cos(k_y) \\
\end{pmatrix}
\Psi_{(k_m,k_y,l^\prime)} \notag\\
&\quad +\Psi_{(k_m,k_y,l)}^\dagger e^{i k_l (l-l^\prime)/\sqrt{2}}\cos\left(\frac{k_l}{\sqrt{2}}\right)
\begin{pmatrix}
 -4 t \cos\left(\frac{k_m}{\sqrt{2}}\right) & i \lambda  \sin\left(\frac{k_m}{\sqrt{2}}\right) & -i \Delta^t \sin\left(\frac{k_m}{\sqrt{2}}\right) & 0 \\
 -i \lambda  \sin\left(\frac{k_m}{\sqrt{2}}\right) & -4 t \cos\left(\frac{k_m}{\sqrt{2}}\right) & 0 & i \Delta^t \sin\left(\frac{k_m}{\sqrt{2}}\right) \\
 i \Delta^t \sin\left(\frac{k_m}{\sqrt{2}}\right) & 0 & 4 t \cos\left(\frac{k_m}{\sqrt{2}}\right) & i \lambda  \sin\left(\frac{k_m}{\sqrt{2}}\right) \\
 0 & -i \Delta^t \sin\left(\frac{k_m}{\sqrt{2}}\right) & -i \lambda 
   \sin\left(\frac{k_m}{\sqrt{2}}\right) & 4 t \cos\left(\frac{k_m}{\sqrt{2}}\right) 
\end{pmatrix}
\Psi_{(k_m,k_y,l^\prime)} \notag \\
&\quad +\Psi_{(k_m,k_y,l)}^\dagger e^{i k_l (l-l^\prime)/\sqrt{2}}\sin\left(\frac{k_l}{\sqrt{2}}\right)
\begin{pmatrix}
0 & i \lambda  \cos\left(\frac{k_m}{\sqrt{2}}\right) & -i \Delta^t
   \cos\left(\frac{k_m}{\sqrt{2}}\right) & 0 \\
 -i \lambda  \cos\left(\frac{k_m}{\sqrt{2}}\right) & 0 & 0 & i \Delta^t \cos\left(\frac{k_m}{\sqrt{2}}\right) \\
 i \Delta^t \cos\left(\frac{k_m}{\sqrt{2}}\right) & 0 & 0 & i \lambda
    \cos\left(\frac{k_m}{\sqrt{2}}\right) \\
 0 & -i \Delta^t \cos\left(\frac{k_m}{\sqrt{2}}\right) & -i \lambda 
   \cos\left(\frac{k_m}{\sqrt{2}}\right) & 0 \\
\end{pmatrix}
\Psi_{(k_m,k_y,l^\prime)} \notag\\
&= \frac{1}{2} \sum_{k_m, k_y, l} \Psi_{(k_m,k_y,l)}^\dagger
\begin{pmatrix}
 -\mu -2 t \cos(k_y) & \lambda  \sin(k_y) & -\Delta^t \sin(k_y) & \Delta^s\\ 
 \lambda  \sin(k_y) & -\mu -2 t \cos(k_y) & -\Delta^s & \Delta^t \sin(k_y) \\
 -\Delta^t \sin(k_y) & \Delta^s & \mu +2 t \cos(k_y) & -\lambda  \sin(k_y) \\
 -\Delta^s & \Delta^t \sin (k_y) & -\lambda  \sin(k_y) & \mu + 2 t \cos(k_y) \\
\end{pmatrix}
\Psi_{(k_m,k_y,l)} \notag\\
&\quad +\Psi_{(k_m,k_y,l+1)}^\dagger
\begin{pmatrix}
-2 t \cos\left(\frac{k_m}{\sqrt{2}}\right) & \frac{\lambda}{2} e^{i k_m/\sqrt{2}} & -\frac{\Delta^t}{2} e^{i k_m/\sqrt{2}} & 0 \\
 -\frac{\lambda}{2} e^{i k_m/\sqrt{2}} & -2 t \cos\left(\frac{k_m}{\sqrt{2}}\right) & 0 & \frac{\Delta^t}{2} e^{i k_m/\sqrt{2}} \\
 \frac{\Delta^t}{2} e^{i k_m/\sqrt{2}} & 0 & 2 t \cos\left(\frac{k_m}{\sqrt{2}}\right) & \frac{\lambda}{2} e^{i k_m/\sqrt{2}} \\
 0 & -\frac{\Delta^t}{2} e^{i k_m/\sqrt{2}} & -\frac{\lambda}{2} e^{i k_m/\sqrt{2}} & 2 t \cos\left(\frac{k_m}{\sqrt{2}}\right) 
\end{pmatrix}
\Psi_{(k_m,k_y,l)} \notag \\
&\quad +\Psi_{(k_m,k_y,l-1)}^\dagger
\begin{pmatrix}
-2 t \cos\left(\frac{k_m}{\sqrt{2}}\right) & \frac{\lambda}{2} e^{i k_m/\sqrt{2}} & -\frac{\Delta^t}{2} e^{i k_m/\sqrt{2}} & 0 \\
 -\frac{\lambda}{2} e^{i k_m/\sqrt{2}} & -2 t \cos\left(\frac{k_m}{\sqrt{2}}\right) & 0 & \frac{\Delta^t}{2} e^{i k_m/\sqrt{2}} \\
 \frac{\Delta^t}{2} e^{i k_m/\sqrt{2}} & 0 & 2 t \cos\left(\frac{k_m}{\sqrt{2}}\right) & \frac{\lambda}{2} e^{i k_m/\sqrt{2}} \\
 0 & -\frac{\Delta^t}{2} e^{i k_m/\sqrt{2}} & -\frac{\lambda}{2} e^{i k_m/\sqrt{2}} & 2 t \cos\left(\frac{k_m}{\sqrt{2}}\right) 
\end{pmatrix}^\dagger
\Psi_{(k_m,k_y,l)},
\label{eq:Hamiltonian_fieldfree}
\end{align}
\end{widetext}
with the Nambu spinor
\begin{align}
\Psi_{(k_m,k_y,l)} &= (c_{(k_m,k_y,l),\uparrow},c_{(k_m,k_y,l),\downarrow}, \nonumber \\
&\qquad c_{(k_m,k_y,l),\uparrow}^\dagger,c_{(k_m,k_y,l),\downarrow}^\dagger)^\top .
\end{align}
In this equation, the sum over $l$ satisfies open boundary conditions, i.e., it has to run over $l=1,\hdots,L$ in the first term, over $l=1,\hdots,L-1$ in the second term and over $l=2,\hdots,L$ in the third term. Equation~\eqref{eq:Hamiltonian_fieldfree} is the Hamiltonian of the field-free system, which is diagonal in $(k_m, k_y)$. It can thus be written as a $4L \times 4L$ matrix, which can be diagonalized to find the properties of the field-free system, such as the spin polarization.
To introduce a field to a strip on the $l=1$ and $l=L$ surfaces, the Hamiltonian has to be Fourier transformed once more. For setups~1 and~2, we substitute
\begin{equation}
c_{(k_m,k_y,l),\sigma}=\frac{1}{\sqrt{M}}\sum_{m\in \lbrace 1, \hdots, M \rbrace}e^{-i k_m [2m+ (l \text{ mod } 2)]/\sqrt{2}}c_{(m,k_y,l),\sigma},
\end{equation}
and for setups~3 and~4,
\begin{equation}
c_{(k_m,k_y,l),\sigma}=\frac{1}{\sqrt{Y}}\sum_{y\in \lbrace 1, \hdots, Y \rbrace}e^{-i k_y y}c_{(k_m,y,l),\sigma}.
\end{equation}
For the Hamiltonian of setups 1 and 2, the term $+(l\mod2)$ in the exponent of the Fourier transformation accounts for the fact that changes in $m$ are not independent of~$l$. In particular, due to $m=\left\lfloor\frac{x-z}{2}\right\rfloor$, moving from a layer with even coordinate $l$ to the next-neighbor site in the odd-numbered layer $l+1$ can only change the coordinate $m$ to $m+1$ if the step is in the $+\mathbf{\hat{e}}_m$ direction, while $m$ stays the same if the step is in the $-\mathbf{\hat{e}}_m$ direction. On the other hand, if $l$ is odd, moving from $l$ to $l+1$ can only change $m\rightarrow m-1$ for a step in the $-\mathbf{\hat{e}}_m$ direction, while a step in the $+\mathbf{\hat{e}}_m$ direction leaves $m$ unchanged.

The boundary conditions in the in-plane direction orthogonal to the strip are periodic, i.e., $m=1$ is equivalent to $m=M$ and $y=1$ is equivalent to $y=Y$.
Thus, we get the Hamiltonian $H=\sum_{k_y}\Psi_{k_y}^\dagger\mathcal{H}(k_y)\Psi_{k_y}$ for setups~1 and~2 and $H=\sum_{k_m}\Psi_{k_m}^\dagger\mathcal{H}(k_m)\Psi_{k_m}$ for setups~3 and~4 with the spinors 
\begin{align}
\Psi_{k_y}&=(\Phi_{m=1,k_y,l=1}, \Phi_{m=2,k_y,l=1}, \Phi_{m=3,k_y,l=1},\dots \notag \\
&\quad~~~\Phi_{m=1,k_y,l=2}, \Phi_{m=2,k_y,l=2},\dots, \dots)
\end{align}
and\begin{align}
\Psi_{k_m}&=(\Phi_{k_m,y=1,l=1}, \Phi_{k_m,y=2,l=1}, \Phi_{k_m,y=3,l=1},\dots \notag \\& \quad~~~\Phi_{k_m,y=1,l=2}, \Phi_{k_m,y=2,l=2},\dots,\dots)
\end{align} and the BdG matrices $\mathcal{H}(k_y)$ and  $\mathcal{H}(k_m)$, respectively. We can then add $4\times4$ blocks
\begin{equation}
\hat{h}=\begin{pmatrix}\mathbf{h} \cdot\boldsymbol{\hat{\sigma}}&0\\0&-(\mathbf{h} \cdot \boldsymbol{\hat{\sigma}})^\top\end{pmatrix}
\end{equation}
to all sites $m$ (in setups~1 and~2) or $y$ (in setups~3 and~4) that belong to the exchange-field strip in the layer $l_\text{field}$, in which the exchange field is applied. For setups~1 and~2, this leads to the BdG Hamiltonian
\begin{widetext}
\begin{equation}\label{eq:H_matrix_set12}
\mathcal{H}(k_y)= 
\begin{pmatrix}
                    &\mbox{\tiny $l_2=1$}&\mbox{\tiny $l_2=2$}&\mbox{\tiny $l_2=3$}& \cdots      &\mbox{\tiny $l_2=L$} \\
\mbox{\tiny $l_1=1$}& D_1(k_y)           & B_1^\dagger(k_y)   & 0                  & \cdots      & 0                   \\
\mbox{\tiny $l_2=2$}& B_1(k_y)           & D_2(k_y)           & B_2^\dagger(k_y)   & \ddots      & \vdots              \\
\mbox{\tiny $l_2=3$}& 0                  & B_2(k_y)           &  D_3(k_y)          & \ddots      & 0                   \\
\vdots              & \vdots             & \ddots             & \ddots             & \ddots      & B_{L-1}^\dagger(k_y)\\
\mbox{\tiny $l_2=L$}& 0                  & \cdots             & 0                  & B_{L-1}(k_y)& D_{L}(k_y)          \end{pmatrix},
\end{equation}
with the diagonal $M{\times}M$ blocks all equal to
\begin{equation}\label{eq:Hamiltonian_set12_block}
\begin{array}{lccccccccl}
                     &\mbox{\tiny $m=1$}&\dots&\mbox{\tiny{$m=m_{min}$}}                            &\dots&\mbox{\tiny $m=m_{max}$}                            &\mbox{\tiny $m=m_{max}+1$}&\dots &\mbox{\tiny $m=M$}&\\
D_l(k_y)=\text{diag}(& \hat{d}(k_y),    &\dots&\hat{d}(k_y)+\delta_{l,l_\text{field}}\hat{h},&\dots&\hat{d}(k_y)+\delta_{l,l_\text{field}}\hat{h}&\hat{d}(k_y),             &\dots &\hat{d}(k_y)&), 
\end{array}
\end{equation}
with
\begin{equation}
\hat{d}(k_y)=\begin{pmatrix}
-2t \cos k_y -\mu             & -\lambda \sin k_y         & -\Delta^y \sin k_y         & \Delta^s          \\
-\lambda \sin k_y             & -2t \cos k_y -\mu         & -\Delta^s                  & \Delta^y \sin k_y \\
-(\Delta^y)^* \sin k_y    &-(\Delta^s)^*              & 2t \cos k_y + \mu          & -\lambda \sin k_y \\
(\Delta^s)^*              & (\Delta^y)^* \sin k_y & -\lambda \sin k_y          & 2t \cos k_y + \mu
\end{pmatrix},
\end{equation}
and the off-diagonal blocks
\begin{equation}
B_l(k_y)=\begin{cases}
\begin{pmatrix}
                      &\mbox{\tiny $m_2=1$}&\mbox{\tiny $m_2=2$}&\ldots&\mbox{\tiny $m_2=M$} \\
\mbox{\tiny $m_1=1$}  & \hat{b}_x          & \hat{b}_z          & 0    & 0                  \\
\mbox{\tiny $m_1=2$}  & 0                  & \ddots             &\ddots& 0                  \\
\vdots                & 0                  & \ddots             &\ddots& \hat{b}_z          \\
\mbox{\tiny $m_1=M$}  & \hat{b}_z          & 0                  & 0    & \hat{b}_x          \\
\end{pmatrix}& \text{for $l$ even,}\\
\begin{pmatrix}
                      &\mbox{\tiny $m_2=1$}&\mbox{\tiny $m_2=2$}& \cdots   &\mbox{\tiny $m_2=M$}\\
\mbox{\tiny $m_1=1$}  & \hat{b}_z          & 0                  & 0        & \hat{b}_x          \\
\mbox{\tiny $m_1=2$}  & \hat{b}_x          & \ddots             & \ddots   & 0                  \\
\vdots                & 0                  & \ddots             & \ddots   & 0                  \\
\mbox{\tiny $m_1=M$}  & 0                  & 0                  & \hat{b}_x& \hat{b}_z          \\
\end{pmatrix}& \text{for $l$ odd,}
           \end{cases}
\end{equation}
\end{widetext}
where
\begin{equation}
\hat{b}_x=\begin{pmatrix}
-t                          & \frac{\lambda}{2}           & \frac{\Delta^x}{2}& 0                  \\
-\frac{\lambda}{2}          & -t                          & 0                 & \frac{\Delta^x}{2} \\
\frac{-(\Delta^x)^*}{2}     & 0                           & t                 & -\frac{\lambda}{2} \\
0                           & \frac{-(\Delta^x)^*}{2}     & \frac{\lambda}{2} & t
\end{pmatrix}
\end{equation}
and
\begin{equation}
\hat{b}_z=
\begin{pmatrix}
-t & 0  & 0  & 0  \\
0  & -t & 0  & 0  \\
0  & 0  & +t & 0  \\
0  & 0  & 0  & +t
\end{pmatrix}.
\end{equation}
For setups 3 and 4, the Hamiltonian matrix is computed analogously, which leads to
\begin{equation}
\mathcal{H}(k_m)=\begin{pmatrix}
                     & \mbox{\tiny $l_2=1$}     & \mbox{\tiny $l_2=2$}     & \cdots                       &\mbox{\tiny $l_2=L$}  \\
\mbox{\tiny $l_1=1$} & \tilde{D}_1              & \tilde{B}^\dagger(k_m)   & 0                            & 0                    \\
\mbox{\tiny $l_2=2$} & \tilde{B}(k_m)           & \ddots                   & \ddots                       & 0                     \\
\vdots               & 0                        & \ddots                   & \ddots                       & \tilde{B}^\dagger(k_m)\\
\mbox{\tiny $l_2=L$} & 0                        & 0                        & \tilde{B}(k_m)               & \tilde{D}_{L}         \\
\end{pmatrix},
\end{equation}
with the $Y{\times}Y$ off-diagonal blocks
\begin{equation}
\tilde{B}(k_m)=\begin{pmatrix}
                     & \mbox{\tiny $y_2=1$} & \ldots & \mbox{\tiny $y_2=Y$} \\
\mbox{\tiny $y_1=1$} & \hat{b}_m(k_m)       & 0      & 0                    \\
\vdots               & 0                    & \ddots & 0                    \\
\mbox{\tiny $y_1=Y$} & 0                    & 0      & \hat{b}_m(k_m)       \\
\end{pmatrix}
\end{equation}
and the diagonal blocks
\begin{widetext}
\begin{equation}\label{eq:Hamiltonian_set34_block}
\tilde{D}_l=\begin{pmatrix}
                           &\mbox{\tiny $y_2=1$}&\mbox{\tiny $y_2=2$}&\mbox{\tiny $y_2=y_{\min}$} &\mbox{\tiny $y_2=y_{\max}$}&\mbox{\tiny $y_2=Y$}\\
\mbox{~~\tiny $y_1=1$~~~}   & \hat{a}            & \hat{b}_y^\dagger  & 0                          & 0                         & \hat{b}_y          \\
\mbox{~~\tiny $y_2=2$~~~}   & \hat{b}_y          & \ddots             & \ddots                     & \ddots                    & 0                  \\
\mbox{\tiny $y_1=y_{\min}$}& 0                  & \ddots             & \hat{a}+\delta_{l, l_\text{field}}\hat{h}& \ddots                     & 0                  \\
\mbox{\tiny $y_1=y_{\max}$}& 0                  & \ddots             & \ddots                      &\hat{a}+\delta_{l,l_\text{field}}\hat{h}& \hat{b}_y^\dagger  \\
\mbox{\tiny ~~$y_1=Y$~~~}   & \hat{b}_y^\dagger  & 0                  & 0                          & \hat{b}_y                 & \hat{a}
\end{pmatrix},
\end{equation}
where
\begin{equation}
\hat{b}_m(k_m)=\begin{pmatrix}
-2t\cos\left(\frac{k_m}{\sqrt{2}}\right)               & \frac{\lambda}{2}e^{\frac{-i k_m}{\sqrt{2}}}           & \frac{\Delta^x}{2}e^{\frac{-i k_m}{\sqrt{2}}}        & 0                                             \\
-\frac{\lambda}{2}e^{\frac{-i k_m}{\sqrt{2}}}          & -2t\cos\left(\frac{k_m}{\sqrt{2}}\right)               & 0                                                    & \frac{\Delta^x}{2}e^{\frac{-i k_m}{\sqrt{2}}} \\
-\frac{(\Delta^x)^*}{2}e^{\frac{-i k_m}{\sqrt{2}}} & 0                                                      & 2t\cos\left(\frac{k_m}{\sqrt{2}}\right)              & -\frac{\lambda}{2}e^{\frac{-i k_m}{\sqrt{2}}} \\
0                                                      & -\frac{(\Delta^x)^*}{2}e^{\frac{-i k_m}{\sqrt{2}}} & \frac{\lambda}{2}e^{\frac{-i k_m}{\sqrt{2}}}         & 2t\cos\left(\frac{k_m}{\sqrt{2}}\right)
\end{pmatrix}
\end{equation}
and
\begin{equation}
\hat{b}_y=\begin{pmatrix}
-t                           & \frac{-i\lambda}{2}         & \frac{-i\Delta^y}{2}& 0                   \\
\frac{-i\lambda}{2}          & -t                          & 0                   & \frac{i\Delta^y}{2} \\
\frac{-i(\Delta^y)^*}{2} & 0                           & t                   & \frac{-i\lambda}{2} \\
0                            & \frac{i(\Delta^y)^*}{2} & \frac{-i\lambda}{2} & t
\end{pmatrix}.
\end{equation}
\end{widetext}

\section{Oscillating spin polarization for setup 4 at the \texorpdfstring{$l=L$}{l=L} surface}
\label{sec:Oscillating m-spin polarization for setup 4 on the l=L surface}

As seen in Sec.~\ref{sec:First-order perturbation theory and spin polarization of the field-free system}, the $m$ component of the spin polarization at the $l=L$ surface is much smaller than at the $l=1$ surface because while the $x$ component of the spin polarization is symmetric when switching from $l$ to $L+1-l$, the $z$ component is antisymmetric. Thus, while $s^x$ and $s^z$ add up to a large $m$ polarization at the $l=1$ surface, they nearly cancel for $l=L$. Indeed, the ASOC vector $\mathbf{l_k} = \mathbf{\hat{x}} \sin k_y - \mathbf{\hat{y}} \sin k_x$ of our model is invariant under a transformation that takes $(k_m,k_y,l)$ to $({-}k_m,k_y,L+1-l)$ and rotates every spin by $\pi$ around the $x$ axis. Due to the relation of the rotation in spinor space and the real-space transformation, this is not a physically allowed transformation and in particular it is not a symmetry of the $C_{4v}$ point group because none of the elements of this group can change the $z$ component. It should instead be interpreted as an artifact of the simple form of the chosen $\mathbf{l_k}$ and would disappear if higher-order symmetry-allowed terms were taken into account. However, these terms are expected to be small so that the following arguments are still a good approximation.

To explain the behavior of the $m$ component of the spin polarization at the $l=L$ surface for setup 4, we give a perturbative argument. As for all $k_y$ the field-free states at $+k_y$ and $-k_y$ are degenerate, the energy shift for a field $\mathbf{h}$ has to start from a superposition of these states, so that
\begin{equation} \label{eq:appendix2_DeltaE}
\Delta E(\pm k_y, k_m) = \mathbf{h} \cdot \langle \mathbf{\hat{s}}_{l}\rangle_{c_1|k_m,k_y,\nu\rangle+ c_2|k_m,-k_y,\nu\rangle},
\end{equation}
where $c_1,c_2\in \mathbb{C}$ are coefficients. For every pair $(k_y, k_m)$, we choose a basis $|{\uparrow}\rangle_l \equiv (1~0)^\top$ and $|{\downarrow}\rangle_l \equiv (0~1)^\top$ of spins pointing in the $+z$ and $-z$ direction, respectively, and consider only unperturbed flat-band surface states, i.e., states with $\nu=1$. Thus, we can parametrize
\begin{equation}
|k_m,k_y\rangle_l=\begin{pmatrix}r_{\uparrow,l}  e^{i \phi_l} \\ r_{\downarrow,l}\end{pmatrix} ,
\end{equation}
with $r_{\uparrow,l}, r_{\downarrow,l} \in [0,1]$ and $\phi_l\in [-\pi,\pi)$. We can express the spin polarizations $s^x_l=\langle\hat{s}^x_l\rangle_{|k_m,k_y\rangle}$, $s^y_l=\langle\hat{s}^y_l\rangle_{|k_m,k_y\rangle}$, and $s^z_l=\langle\hat{s}^z_l\rangle_{|k_m,k_y\rangle}$ as well as the weight $p_l=\langle\hat{\sigma}^0_l\rangle_{|k_m,k_y\rangle}$ of the state localized in the layer $l$ in terms of these parameters:
\begin{align}
s^x_l&=2 r_{\uparrow,l} r_{\downarrow,l} \cos\phi_l,\\
s^y_l&=2 r_{\uparrow,l} r_{\downarrow,l} \sin\phi_l,\\
s^z_l&=r_{\uparrow,l}^2 -r_{\downarrow,l}^2,\\
p_l&=r_{\uparrow,l}^2+r_{\downarrow,l}^2 ,
\end{align}
which can be solved for $r_{\uparrow,l}$, $r_{\downarrow,l}$, and $\phi_l$, leading to
\begin{align}
r_{\uparrow\downarrow,l}&=\sqrt{\frac{p_l\pm s^z_l}{2}}, \\
  \phi_l &= \begin{cases}
    \arccos\left(\frac{s^x_l}{\sqrt{p_l^2-(s^z_l)^2}}\right)
      & \text{for $s^x_l>0$} \\*[2ex]
    \pi-\arccos\left(\frac{-s^x_l}{\sqrt{p_l^2-(s^z_l)^2}}\right)
      &  \text{for $s^x_l<0$} .
  \end{cases}
\end{align}
Due to the antisymmetry of $s^x$ and $s^z$  with respect to $k_y$, a sign change of $k_y$ thus replaces $\phi_l$ by  $\pi-\phi_l$ and interchanges $r_\uparrow$ and $r_\downarrow$. This means that the superposition $|\Psi\rangle \equiv c_1|k_m,k_y,\nu\rangle+ c_2|k_m,-k_y,\nu\rangle$ can be written as 
\begin{align}
|\Psi\rangle &=
  c\  e^{i \varphi}|k_m,k_y\rangle+\sqrt{1- c^2}\, |k_m,-k_y\rangle \nonumber \\
&= c\ e^{i \varphi}e^{i k_y y}\begin{pmatrix}
r_{\uparrow,l} e^{i \phi_l}\\ r_{\downarrow,l}\end{pmatrix} \notag \\
&\quad{}+ \sqrt{1- c^2}\, e^{-ik_y y}\begin{pmatrix}r_{\downarrow,l} e^{i (\pi-\phi)}\\ r_{\uparrow,l}\end{pmatrix} ,
\end{align}
with $c\in [0,1]$ and $\varphi\in[-\pi,\pi)$. Substituting this state into $\langle\hat{\mathbf{s}}_l\rangle_{|\Psi\rangle}$ and transforming to $s^m=(s^x-s^z)/\sqrt{2}$ gives 
\begin{align}
s^m_1(y) &= 4 c \sqrt{1-c^2} \cos(2k_y y +\varphi +\phi) (s^x_1+s^z_1)  \notag \\
&\quad{}+ 2 (2 c^2-1) (s^x_1-s^z_1),\\
s^m_L(y) &= -4  c \sqrt{1- c^2} \cos(2k_y y + \varphi +\phi) (s^x_1-s^z_1)   \notag \\
&\quad{}- 2 (2 c^2-1) (s^x_1+s^z_1)
\end{align}
as the $m$ component of the spin polarization of the superposition in Eq.\ (\ref{eq:appendix2_DeltaE}), for the two surfaces. The unperturbed state $|\Psi\rangle_0\equiv c_0 e^{i \varphi_0}|k_m,k_y\rangle+\sqrt{1-c_0^2}\, |k_m,-k_y\rangle$ that is closest to the ground state of the perturbed system generically corresponds to a linear combination of the states $|k_m,k_y\rangle$ and $|k_m,-k_y\rangle$ with a coefficient $c=c_0\neq 0$. Thus, for $s^x_1\approx -s^z_1$, which is the case for our model system, perturbation theory starts from a state with finite and spatially almost constant spin polarization  $s^m_1(y)$ at the $l=1$ surface and from a state with spatially strongly oscillating but on average almost vanishing spin polarization $s^m_L(y)$ at the $l=L$ surface. In the latter case, this means that the effects of an exchange field $\mathbf{h}=h \hat{\mathbf{e}}_m$ in regions with positive and negative spin polarizations cancel everywhere except if they lie on different sides of the boundary between the exchange-field and the field-free strip. Thus, the only states which are linearly shifted away from zero energy in setup 4 at the $l=L$ surface are localized at these boundaries.

\bibliography{lapp}

\end{document}